\documentclass[aps,superscriptaddress,showpacs]{revtex4} 

\input epsf

\begin{document}

\title{Influence of mass difference on dynamic properties of isotope mixtures}
\author{N. Kiriushcheva}
\email{nkiriusc@uwo.ca}
\author{S.V. Kuzmin}
\affiliation{Department of Applied Mathematics, University of Western
Ontario, London, Ontario N6A~5B7, Canada}

\date{\today}

\begin{abstract}
We present the results of a Molecular Dynamics computer simulations of a two
component isotope mixture of Lennard-Jones particles, monodisperse in size
but different in masses, at a fixed average density and temperature. We
study changes in properties that result from mass heterogeneity, by
measuring the pair distribution function, diffusion coefficient, velocity
autocorrelation function, non-Gaussian and isotope effect parameters, as
functions of the degree of mass difference. Our results show that if static
properties are not influenced by a variation of mass variation, the dynamic
properties are significantly affected and even exhibit the presence of
``critical'' values in the mass difference. We also demonstrate that our model
gives a simple contra example to a recently proposed a universal scaling law 
for atomic
diffusion in condensed matter [M. Dzugudov, Nature, {\bf 381}, 137 (1996)].
\end{abstract}

\maketitle

\section{Introduction}

The aim of this paper is to study, using Molecular Dynamics (MD) computer
simulations, the effect of having a mass difference in particles on 
properties of
liquids. The influence of mass polydispersity on dynamical properties when
simulating the dynamics of realistic system having size polydispersity was
recently demonstrated by Poole and one of the authors (N.K.) \cite{KP}.
Computer simulation methods, MD in particular, gives unique possibility to
isolate and study the role of a single parameter for systems of any
complexity that is much more difficult to accomplish (or even impossible)
in an actual experiment. 

Here we consider a model that allows us to isolate completely the effect of 
only changing the masses of the particles by keeping all other 
characteristics of system unchanged. Our model is the simplest possible 
isotope system: a two-component mixture of isotopes, particles of the same 
size but different masses, consisting of
50\% particles $A$ with mass $m_A=m_0-\Delta m$ and 50\% particles $B$ with
mass $m_B=m_0+\Delta m$, so that no matter how we change $\Delta m$, the
average mass of particles, total mass and average mass and number densities
of a system remain the same. In this sense, our model differs from these 
studied previously. For example, Ebbsj\"o {\it et al. }in \cite{Ebbsjo} and
Bearman and Jolly in \cite{Bearman} studied isotope systems in which one
species had a fixed mass and the mass of another species varied. 
Consequently, this affected
the average mass and mass density. Other authors \cite
{Herman,Easteal,Ould-Kaddour,Nuevo,Ould2} considered the diffusion of solute
particles in the limit of infinite dilution in a solvent.

We perform the equilibrium MD simulation of a 3-dimensional system of $N=4000
$ particles interacting via the shifted-force Lennard-Jones (LJ) potential 
\cite{Allen}, a modification of the standard LJ potential 
\begin{equation}
\label{lj}V(r)=4\varepsilon \left[ \left( \frac \sigma r\right) ^{12}-\left(
\frac \sigma r\right) ^6\right] ,
\end{equation}
where $\varepsilon $ characterizes the strength of the pair interaction and $%
\sigma $ describes the particle diameter. Both $\varepsilon $ and $\sigma $
are constant for all particle pairs. In the shifted-force LJ interaction,
the LJ potential and force are modified so as to go to zero continuously at $%
r=2.5\sigma $, and interactions beyond $2.5\sigma $ are ignored. We chose
the Lennard-Jones potential because it is very popular in MD studies and has 
proved to be a good approximation, for example, for liquid argon. In
addition, there is a large amount of data, both theoretical and
experimental, available for comparison \cite{MarchTosi,Nicolas,Rahman,Rice}. 

In our work we use reduced units. Energy is expressed in units of $%
\varepsilon $, mass $m$ in units of $m_0$, length in units of $\sigma $, the
number density of particles $\rho $ in units of $\sigma ^{-3}$, and
temperature $T$ in units of $\varepsilon /k$, where $k$ is Boltzmann's
constant. Time $t$ is expressed in units of $\sqrt{m_0\sigma ^2/\varepsilon }
$. In these units the time step used for integrating the particle equations
of motion is $0.01$. The initial distribution of velocities was Maxwellian.

After equilibration, all quantities are evaluated in the
microcanonical ensemble (a constant NVE). We present data for $\rho =0.75$ and 
$T=0.66$ that is not far from a triple point of one component fluid ($\rho
_t=0.85,T_t=0.76$ \cite{BoonYip}) where the cooperative motion of liquid
particles is more prominent. We conduct simulations for different $\Delta
m$ from $0$ to $0.7$ with the step $0.1$.

\section{Pair distribution function, diffusion coefficient, and velocity
autocorrelation function}

\begin{figure}
\hbox to\hsize{\epsfxsize=1.0\hsize\hfil\epsfbox{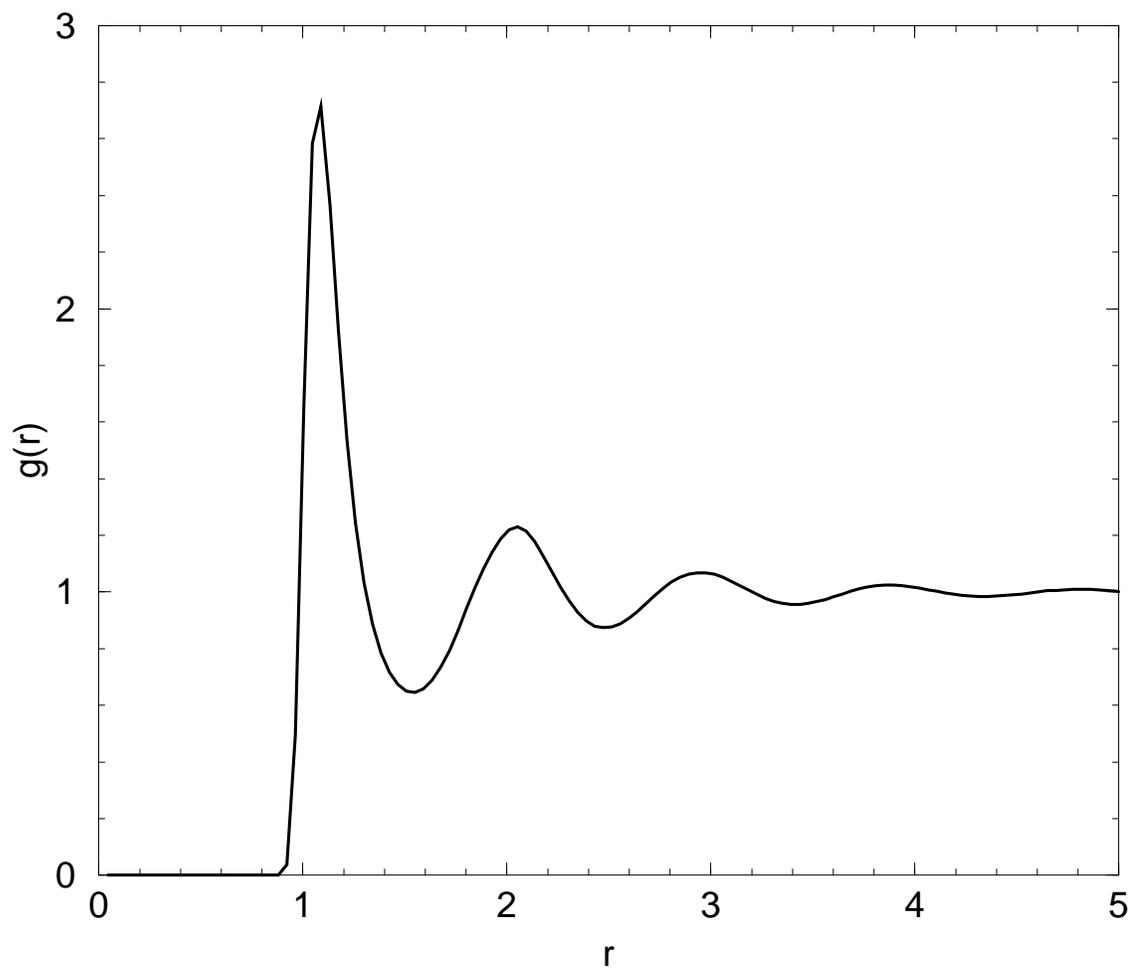}\hfil}
\caption{The pair distribution function $g(r)$ of all species.}
\label{g(r)}
\end{figure}

The pair distribution function $g(r)$ that characterizes the average liquid
structure \cite{HansenMcDonald} is shown in Fig.~\ref{g(r)}. We calculate $%
g(r)$ for systems with different values of $\Delta m$ and for each species $%
A $ and $B$ of these systems. We find that the pair distribution function $%
g(r) $ is the same in all cases, confirming that the static properties of
the system do not depend on $\Delta m$ \cite{HansenMcDonald}.

\begin{figure}
\hbox to\hsize{\epsfxsize=1.0\hsize\hfil\epsfbox{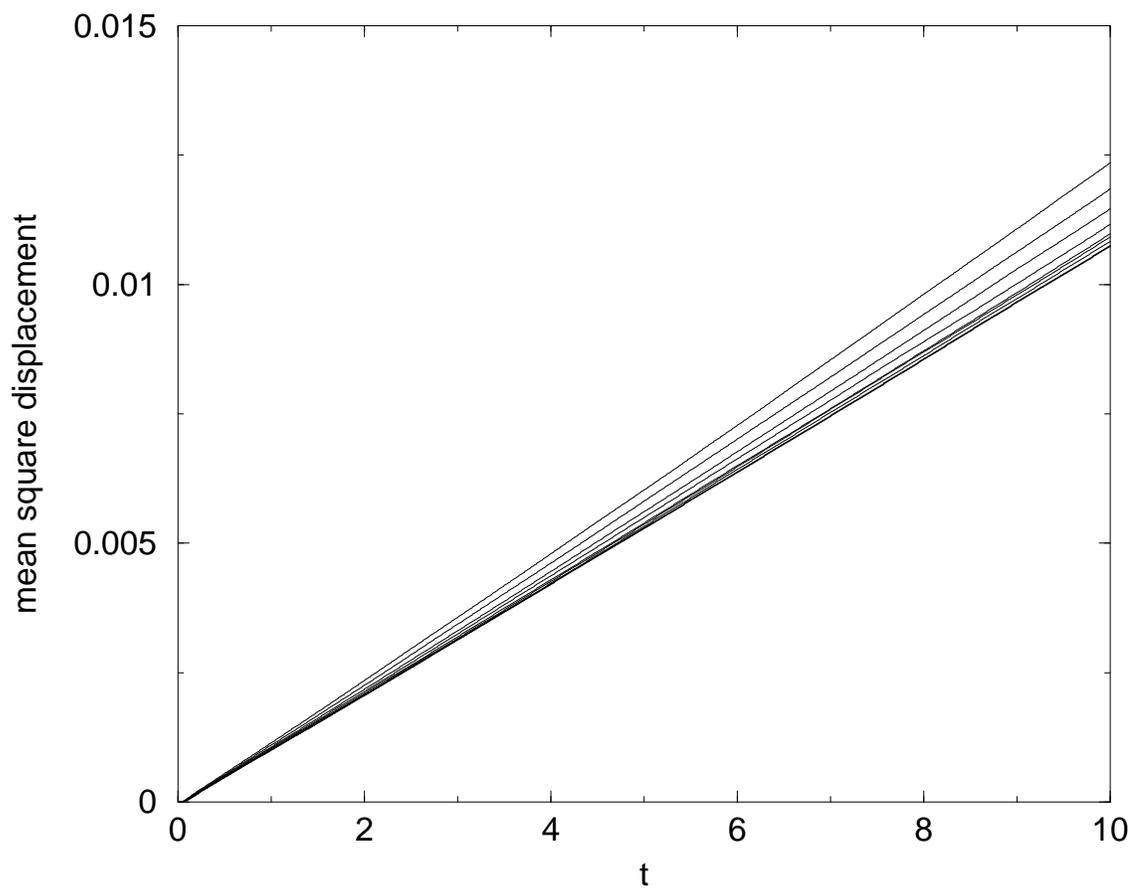}\hfil}
\caption{The mean square displacement versus $t$ for increasing $\Delta m$ 
(from 
the bottom to the top). }
\label{msqdis}
\end{figure}

\begin{figure}
\hbox to\hsize{\epsfxsize=1.0\hsize\hfil\epsfbox{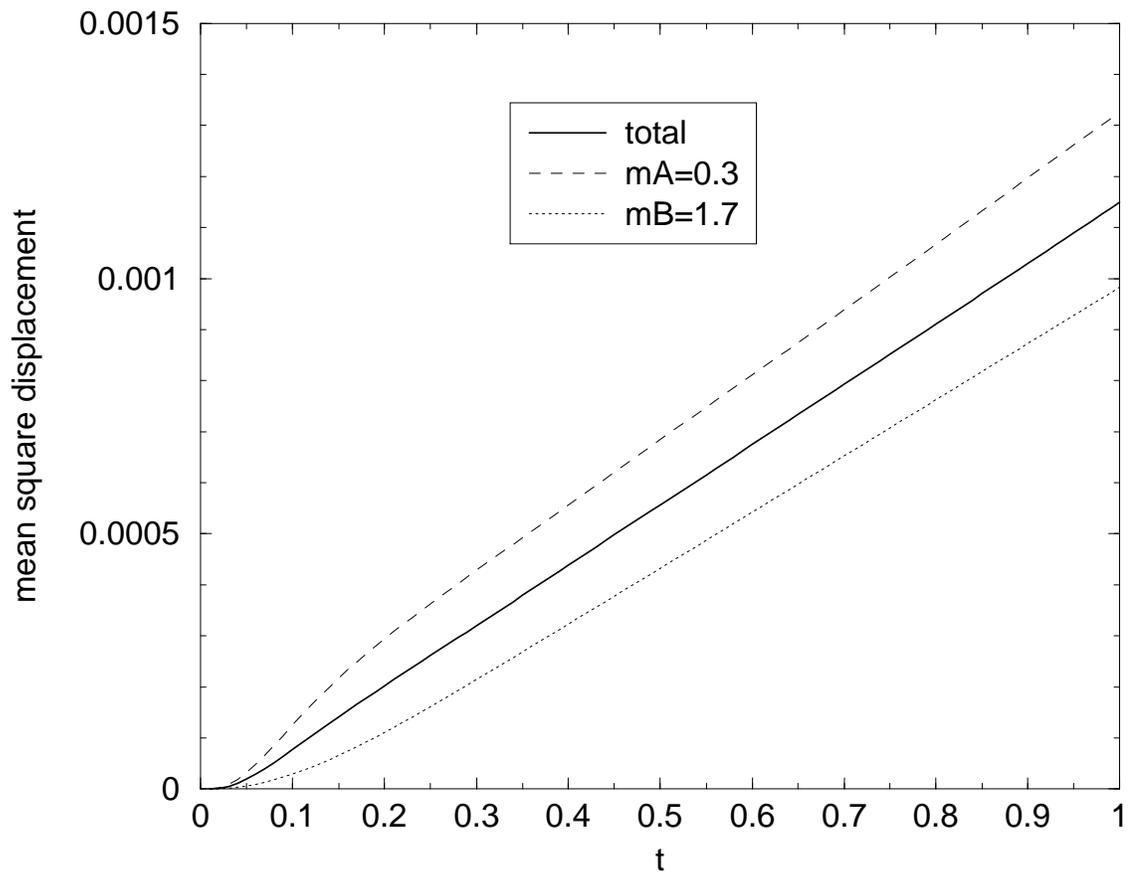}\hfil}
\caption{The mean square displacement versus $t$ for lighter, heavier 
particles and the total 
for $\Delta m = 0.7$. }
\label{msqdis3}
\end{figure}

Fig.~\ref{msqdis} presents the dependence of $\langle |{\bf r}(t)-{\bf r}%
(0)|^2\rangle $ (the mean square displacement of a particle) on $t$ for
different $\Delta m$. As $\Delta m$ becomes larger, $\langle |{\bf r}(t)-%
{\bf r}(0)|^2\rangle $ has a steeper slope. In addition, in Fig. \ref
{msqdis3} we plot three graphs that show the mean square displacements of
particles $A$, $B$, and the total $\langle |{\bf r}(t)-{\bf r}(0)|^2\rangle $
for $\Delta m=0.7$, as an example. The curves for particles with mass $%
m_A=m_0-\Delta m$ lie above and for particles with mass $m_B=m_0+\Delta m$
lie below the curve for the total mean square displacement. This
demonstrates the fact that lighter particles not only move faster between
collisions, but also diffuse faster than heavier particles. It is
interesting that for small $t$ the graph of the mean square displacement of
a lighter particle has a steep slope which decreases and becomes constant
for $t>0.15$, whereas the graph of the mean square displacement of a heavier
particle increases for $0<t<0.15$.

We calculate the diffusion coefficient $D$ in two ways: using the Einstein
relation \cite{HansenMcDonald} 
\begin{equation}
\label{2}D_E=\lim _{t\to \infty }\frac{\langle |{\bf r}(t)-{\bf r}%
(0)|^2\rangle }{6t} 
\end{equation}
and using the Kubo formula \cite{HansenMcDonald}

\begin{equation}
\label{3}D_K=\frac 13\int_0^\infty \left\langle {\bf v }(0)\cdot {\bf v }%
(t)\right\rangle dt, 
\end{equation}
where $\left\langle {\bf v }(0)\cdot {\bf v }(t)\right\rangle $ is the
velocity autocorrelation function. Here and below subscripts $E$ and $K$
mean that $D$ calculated using the Einstein formula and using the Kubo
formula.

The results are presented in the Table~\ref{tab1}. $D_{A(B)}$ is the
diffusion coefficient in scaled units for species with mass $m_{A(B)}$, and $%
D$ is the average diffusion coefficient of the mixture. At fixed $\rho $ and 
$T$ the diffusion coefficients $D$, $D_A$, and $D_B$ increase as $\Delta m$
increases.

$D_E$ and $D_K$ are equal within the computational uncertainties;
however we find that calculations using Eq. (\ref{2}) are more accurate. The
same conclusion was made in Ref. \cite{Bernu}. This is because Eq. (\ref{3})
involves integration where negative and positive parts of $\left\langle {\bf %
v}(0)\cdot {\bf v}(t)\right\rangle $ partially cancel out and because of
uncertainties in the long-time behavior of the velocity autocorrelation
function, whereas the graphs of $\langle |{\bf r}(t)-{\bf r}(0)|^2\rangle $
versus $t$ are almost perfect straight lines for $t>0.15$.

\begin{figure}
\hbox to\hsize{\epsfxsize=1.0\hsize\hfil\epsfbox{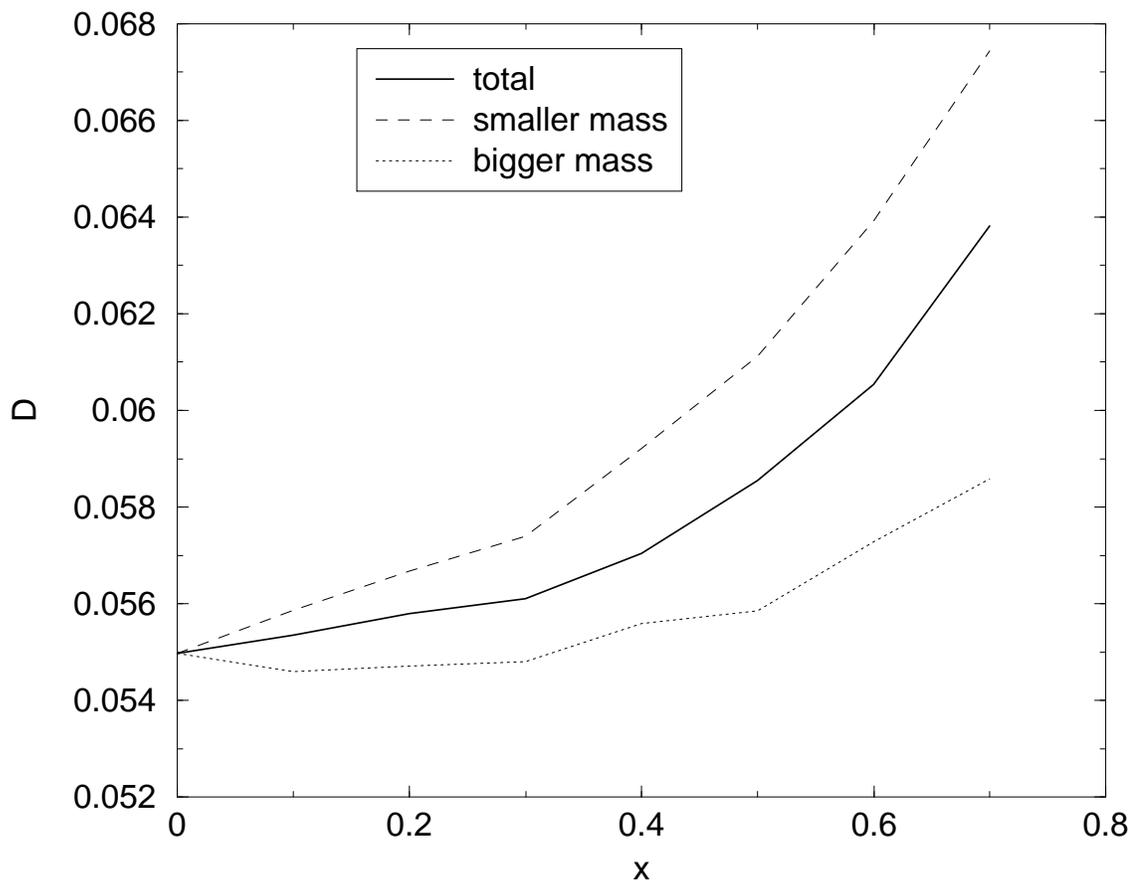}\hfil}
\caption{The diffusion coefficient $D$ versus $x = \Delta m/m_0$ for lighter, 
heavier particles, 
and the total. }
\label{diff}
\end{figure}

From the Table \ref{tab1} we can see that the relationship $D_A>D>D_B$
always holds and the difference among $D_A$, $D$, and $D_B$ increases with an 
increase of $\Delta m$. This is shown in Fig. \ref{diff}. The dependence of
the diffusion coefficient versus $\Delta m$ is not linear.

\begin{figure}
\hbox to\hsize{\epsfxsize=1.0\hsize\hfil\epsfbox{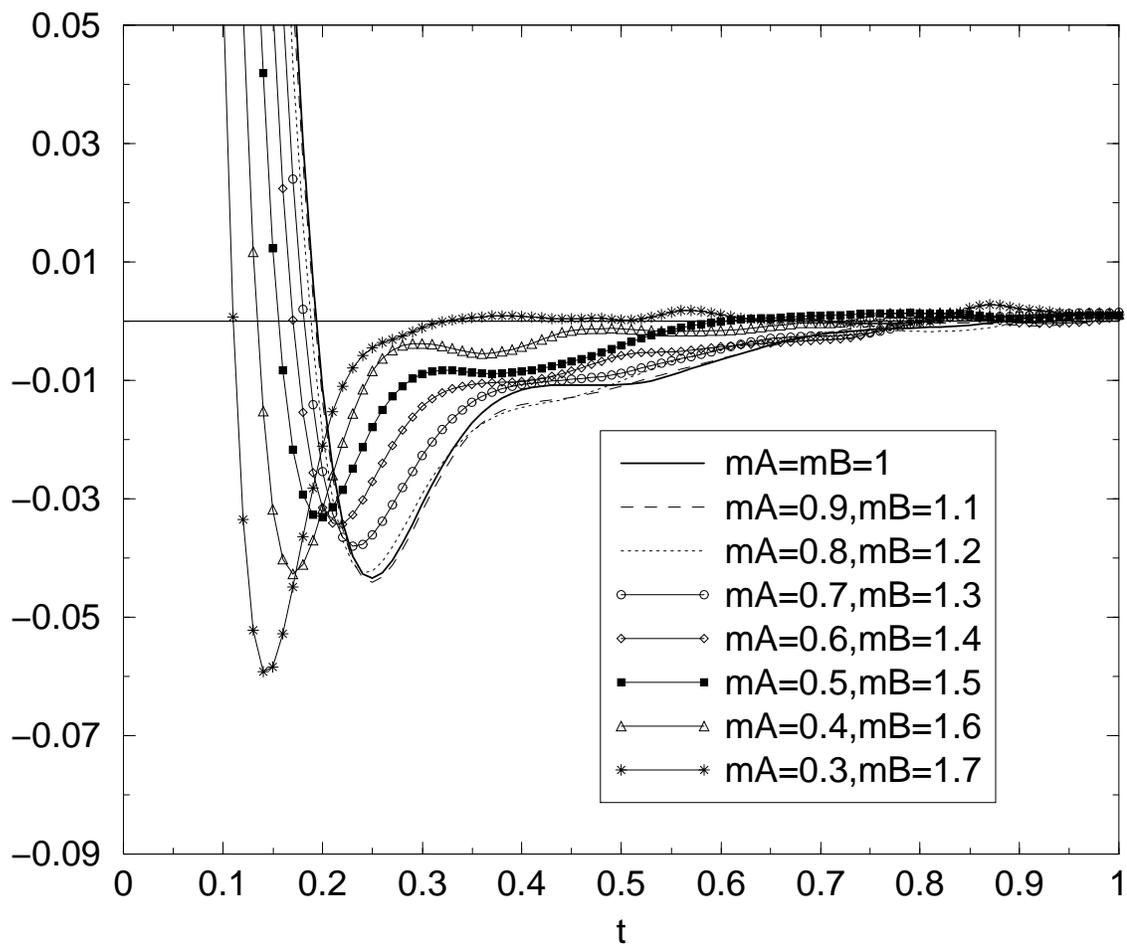}\hfil}
\caption{The velocity autocorrelation function $\psi(t)$ for different 
$\Delta m$.}
\label{vacf}
\end{figure}

Even though the diffusion coefficient does not change too much with a change
of mass (if $\Delta m$ changes by $1\%$, $D$ changes maximum by $5\%$), the
velocity autocorrelation function varies significantly, both in shape and
magnitude as a function of $t$. Fig.~\ref{vacf} shows the dependence on $%
\Delta m$ of the normalized velocity autocorrelation function $\psi (t)$ 
\cite{BoonYip}

\begin{equation}
\label{4}\psi (t)=\frac{\left\langle {\bf v}(0)\cdot {\bf v}(t)\right\rangle 
}{\left\langle |{\bf v}(0)|^2\right\rangle }. 
\end{equation}

\begin{figure}
\hbox to\hsize{\epsfxsize=1.0\hsize\hfil\epsfbox{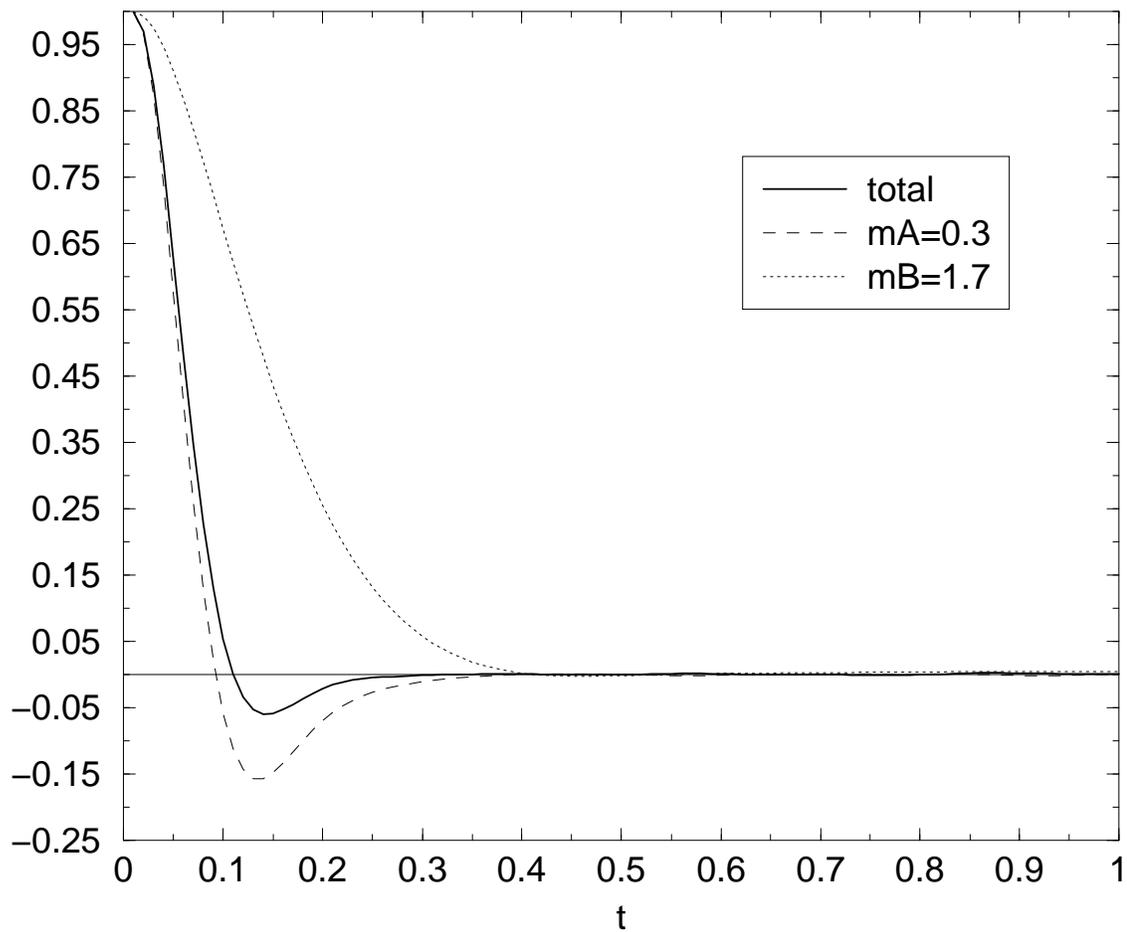}\hfil}
\caption{The velocity autocorrelation function $\psi(t)$ for lighter, 
heavier particles, 
and the total for $\Delta m = 0.7$.}
\label{vacf3}
\end{figure}

It is interesting to observe the behavior of the negative part of $\psi (t)$.
The velocity autocorrelation function, basically, is the projection of a
particle's velocity at time $t$ on the initial velocity of that particle,
averaged over all particles. If $\psi (t)$ becomes negative it means that a
particle after a number of collisions, on average, reverses the direction of
its motion. As we can see from the graph, all of the functions $\psi (t)$ 
have a negative part but the minimum of $\psi (t)$ behaves differently. 
For $\Delta m=0.1$ and $0.2$, the functions $\psi (t)$ are very close 
to $\psi (t)$ with $\Delta m=0$. As $\Delta m$ increases, the minimum shifts 
to earlier times. Its magnitude initially decreases, has a maximum at 
$\Delta m=0.5$, and starts to increase drastically for $\Delta m\geq 0.5$. 
To reveal this difference we plot three curves on Fig. \ref{vacf3} which 
shows the velocity autocorrelation functions of particles $A$, $B$, and the 
total value for $\Delta m=0.7$. The curve
for particles with mass $m_A=m_0-\Delta m$ has a deeper minimum than the
total and the curve for particles with mass $m_B=m_0+\Delta m$ does not have
a negative part at all. For a big difference in the masses $m_A$ and $m_B$,
the heavier particles do not reverse their direction of motion on average,
even in a dense fluid.

\section{Non-Gaussian and isotope effect parameters}

The general non-Gaussian parameter $\alpha _n(t)$ is defined as \cite
{BoonYip}

\begin{equation}
\label{5a}\alpha _n(t)=\frac{\langle r^{2n}(t)\rangle }{c_n\langle
r^2(t)\rangle ^n}-1, 
\end{equation}

where $c_n=[1\cdot 3\cdot 5\cdot \cdot \cdot (2n+1)]/3^n$ and $\langle
r^{2n}(t)\rangle $ are the ensemble average of the $2n^{th}$ power of the
particle displacements after a time $t$ \cite{Rahman}:

\begin{equation}
\label{rdef1}\langle r^{2n}(t)\rangle =\left\langle \frac 1N\sum_{i=1}^N|%
{\bf r}_i(t)-{\bf r}_i(0)|^{2n}\right\rangle . 
\end{equation}

For systems in which motions of particles are uncorrelated (for example, in
an ideal gas) $\alpha _n(t)=0$. The deviation of $\alpha _n(t)$ from zero
serves to quantify the correlation of particle motions at intermediate time
or, as it was shown in \cite{KP}, the heterogeneity in masses for short times.

Here, as in \cite{KP}, we calculate $\alpha _2(t)$:

\begin{equation}
\label{5}\alpha _2(t)=\frac{3\langle r^4(t)\rangle }{5\langle r^2(t)\rangle
^2}-1. 
\end{equation}

\begin{figure}
\hbox to\hsize{\epsfxsize=1.0\hsize\hfil\epsfbox{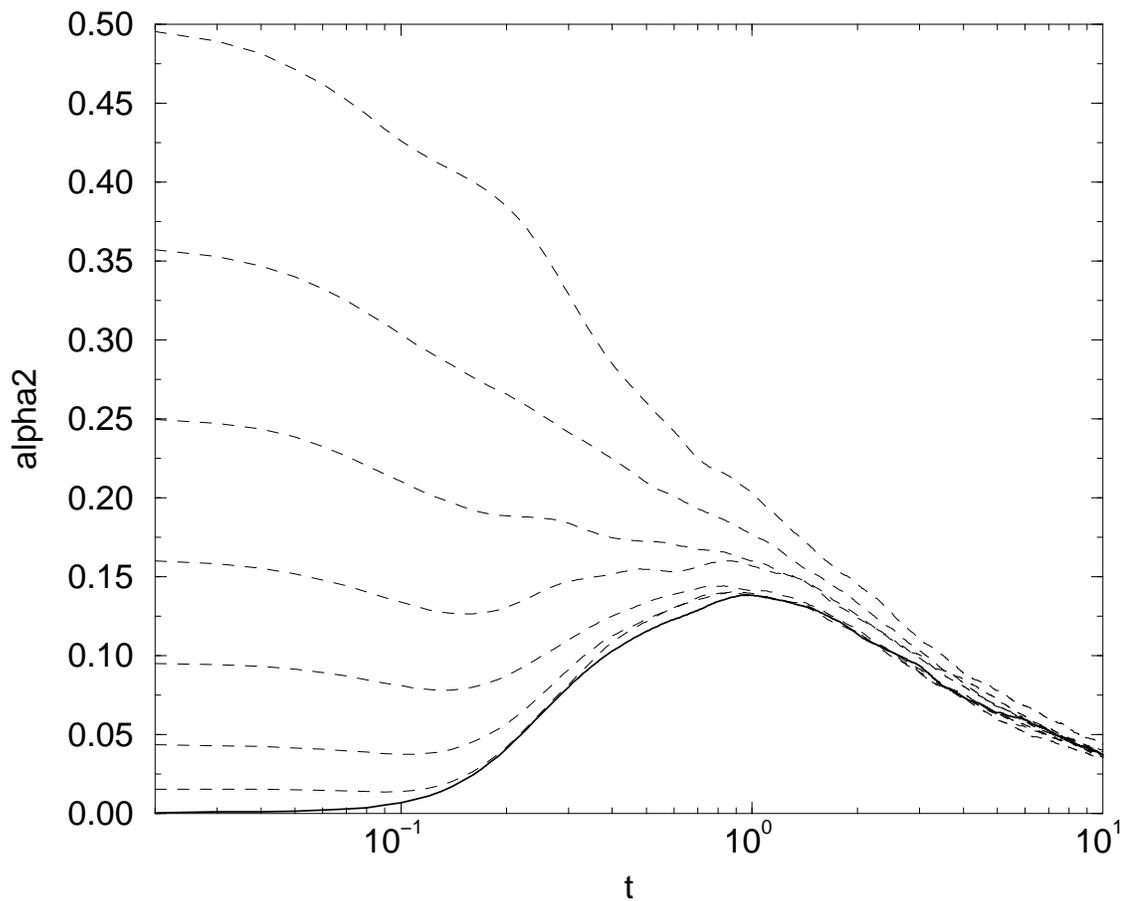}\hfil}
\caption{$\alpha_2(t)$ for $\Delta m = 0, 0.1,...,0.7$ (from the bottom 
to the top).
$\alpha_2^{\circ }$ increases with increase of $\Delta m$.}
\label{alpha2}
\end{figure}

\begin{figure}
\hbox to\hsize{\epsfxsize=1.0\hsize\hfil\epsfbox{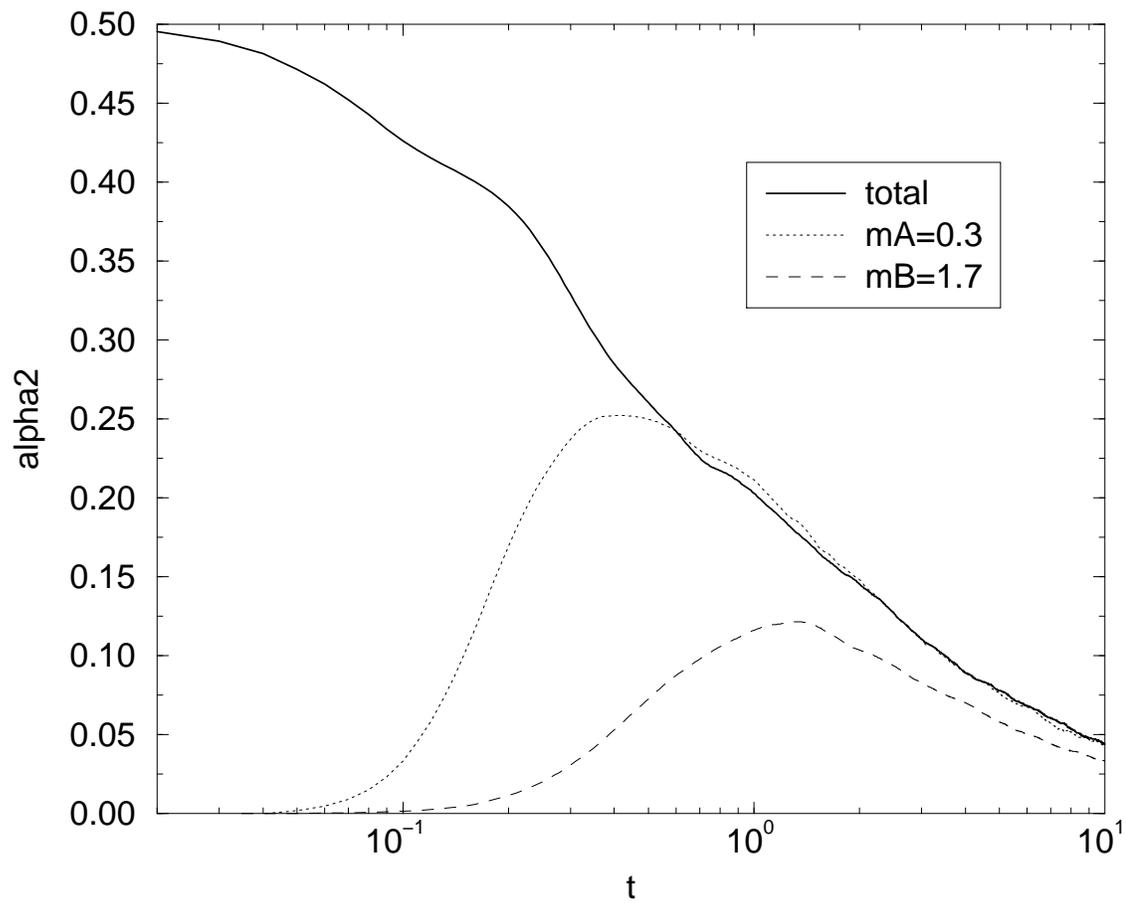}\hfil}
\caption{$\alpha_2(t)$ for lighter, heavier particles, 
and the total for $\Delta m = 0.7$.}
\label{alpha23}
\end{figure}

In Fig. \ref{alpha2} we plot $\alpha _2(t)$ for different values of $\Delta
m $. For small $\Delta m$ the behavior of $\alpha _2(t)$ is similar to that
of a polydisperse system \cite{KP}. For $\Delta m\neq 0$, $\alpha _2$ does
not start from $0$. We observe the characteristic, intermediate-time peak of 
$\alpha _2$, at approximately $t=1$, which increases in magnitude as $\Delta
m $ increases. As in \cite{KP}, the combination of the non-zero, early-time
behavior of $\alpha _2$, and the intermediate-time peak produces a minimum
in $\alpha _2$ at approximately $t=0.12$. However, for larger $\Delta m$,
starting from $\Delta m=0.5$, the minimum disappears, and $\alpha _2(t)$
becomes a monotonically decreasing function. For these $\Delta m$ the values
of $\alpha _2(t)$ at $t\rightarrow 0$ even exceed the intermediate-time
peak. It is interesting to observe how the total non-Gaussian parameter is
related to ones of particles $A$ and $B$. Fig. \ref{alpha23} shows three
curves representing $\alpha _2(t)$ for particles $A$, $B$, and the total 
when $\Delta m=0.7$. Both curves ($A$ and $B$) start from $0$ 
(each component has
particles of the same size). The curve for particles with mass $%
m_A=m_0-\Delta m$ has a maximum that is twice as high and is located at an 
earlier time than the curve for particles with mass $m_B=m_0+\Delta m$. The
difference between these two curve increases when $\Delta m$ becomes larger.
In addition, the curve for the lighter particle, after passing the maximum,
completely coincides with the total $\alpha _2(t)$ and decays as $t^{-1}$,
according to \cite{Yamaguchi}. The lighter particles are more mobile and
give a larger contribution to the motion of the fluid.

As we pointed out in \cite{KP}, the non-Gaussian behavior of $\alpha _2(t)$
shows that the self-part of the van Hove correlation function $G_s({\bf r}%
,t) $

\begin{equation}
\label{Gs}G_s(r,t)=\left\langle \frac 1N\sum_{i=1}^N\delta (r-|{\bf r}_i(t)-%
{\bf r}_i(0)|)\right\rangle . 
\end{equation}
deviates from the Gaussian form. In the limit ${\bf r},t\rightarrow 0$, when
particles move with a velocity ${\bf v}_i={\bf r}_i/t$ as if they were free,
this corresponds to the condition that the distribution of velocities of
particles is non-Maxwellian. The Maxwell-Boltzmann distribution for particle
velocities is \cite{HansenMcDonald}

\begin{equation}
\label{maxw}\phi _i({\bf \upsilon })=\left( \frac{m_i}{2\pi kT}\right)
^{3/2}\exp \left( -\frac{m_i|{\bf \upsilon |}^2}{2kT}\right) . 
\end{equation}

We can see that it depends on the mass of a particle. Even if for each
species the distribution is Maxwellian the total distribution is not

\begin{equation}
\label{maxwt}\phi _{total}({\bf \upsilon })=\frac{N_A}N\left( \frac{m_A}{%
2\pi kT}\right) ^{3/2}\exp \left( -\frac{m_A|{\bf \upsilon |}^2}{2kT}\right)
+\frac{N_B}N\left( \frac{m_B}{2\pi kT}\right) ^{3/2}\exp \left( -\frac{m_B|%
{\bf \upsilon |}^2}{2kT}\right) . 
\end{equation}

\begin{figure}
\hbox to\hsize{\epsfxsize=1.0\hsize\hfil\epsfbox{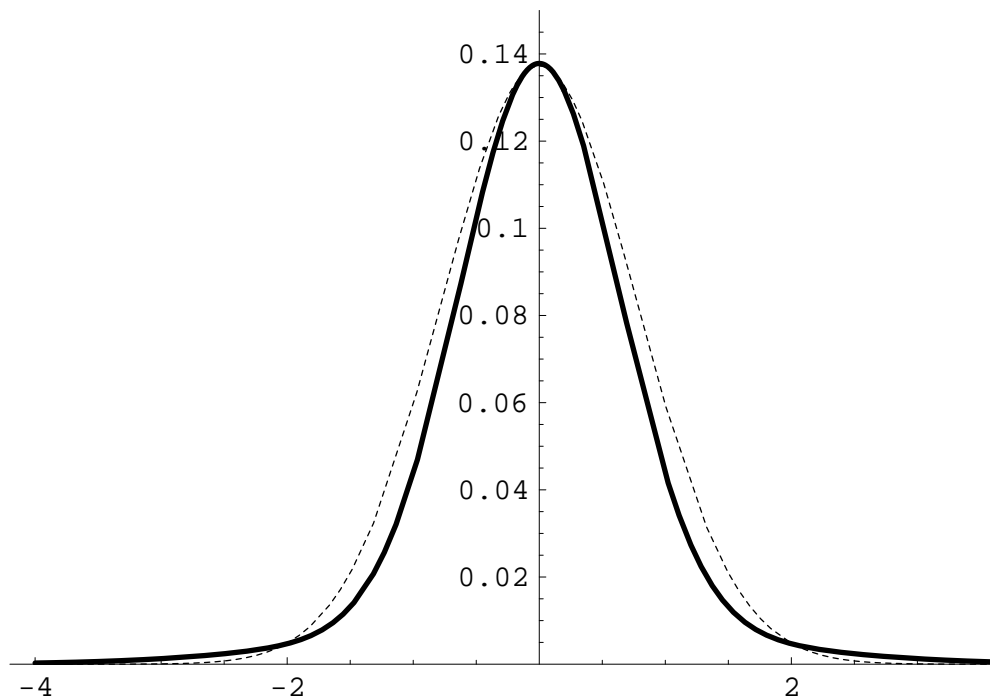}\hfil}
\caption{Distribution of velocities of particles (solid line) and 
the Gaussian fit
(dashed line) for $\Delta m = 0.7$.}
\label{maxvel}
\end{figure}

In Fig. \ref{maxvel} we plot the distribution of velocities of all particles
(solid line) and the Gaussian fit of this curve (dashed line). It is clear
that this distribution deviates from the Gaussian form. It is narrower and
has longer tails for large velocities. This is the contribution of 
particles with
smaller mass. It is interesting that the same non-Gaussian behavior of $G_s(%
{\bf r},t)$ was observed experimentally in \cite{Kasper} for colloid
suspensions. It could be accounted for the polydispersity in the mass of
colloids used in that experiment (which was $\sim 8\%$).

Finally, we can use the formula for the non-Gaussian parameter $\alpha _2$
at $t\rightarrow 0$ for the multicomponent system from \cite{KP}:

\begin{equation}
\label{alpha0}\alpha _2^{\circ }\equiv \alpha _2(t\rightarrow 0)=\frac{%
\sum_{i=1}^ln_i/m_i^2}{\left( \sum_{i=1}^ln_i/m_i\right) ^2}-1, 
\end{equation}
where $n_i=N_i/N$ is the fraction of particles of species $i$. (Here, as in  
\cite{KP}, the superscript ``$\circ $'' indicates the limit $t\to 0$.)

If we have a binary system with two species having  masses $m_A=m_0-\Delta m$ 
and $m_B=m_0+\Delta m$, Eq. (\ref{alpha0}) becomes

\begin{equation}
\label{alpha1}\alpha _2^{\circ }=\frac{\frac{n_A}{(m_0-\Delta m)^2}+\frac{nB%
}{(m_0+\Delta m)^2}}{\left( \frac{n_A}{m_0-\Delta m}+\frac{n_B}{m_0+\Delta m}%
\right) ^2}-1=\frac{4n_An_Bx^2}{(1+(n_A-n_B)x)^2}, 
\end{equation}
where $x\equiv \Delta m/m_0$. In our case $n_A=n_B=1/2$ and

\begin{equation}
\label{alphax2}\alpha _2^{\circ }=x^2. 
\end{equation}

\begin{figure}
\hbox to\hsize{\epsfxsize=1.0\hsize\hfil\epsfbox{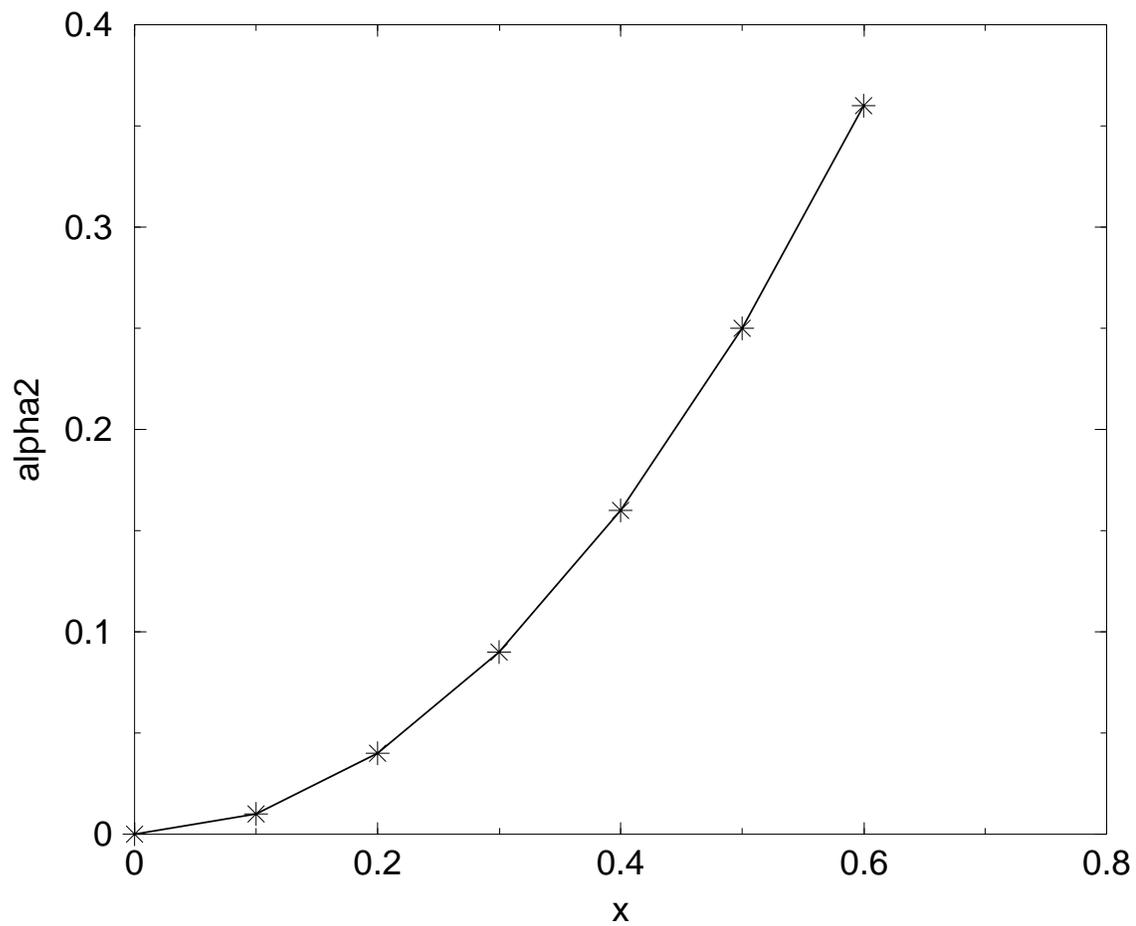}\hfil}
\caption{$\alpha_2$ versus $x = \Delta m /m_0$. Stars represent results of 
computer simulation and the solid line is Eq. \ref{alphax2}. } 
\label{alpha2x}
\end{figure}

We plot $\alpha _2^{\circ }$ versus $\Delta m$ in Fig. \ref{alpha2x},
confirming this analysis.

\begin{figure}
\hbox to\hsize{\epsfxsize=1.0\hsize\hfil\epsfbox{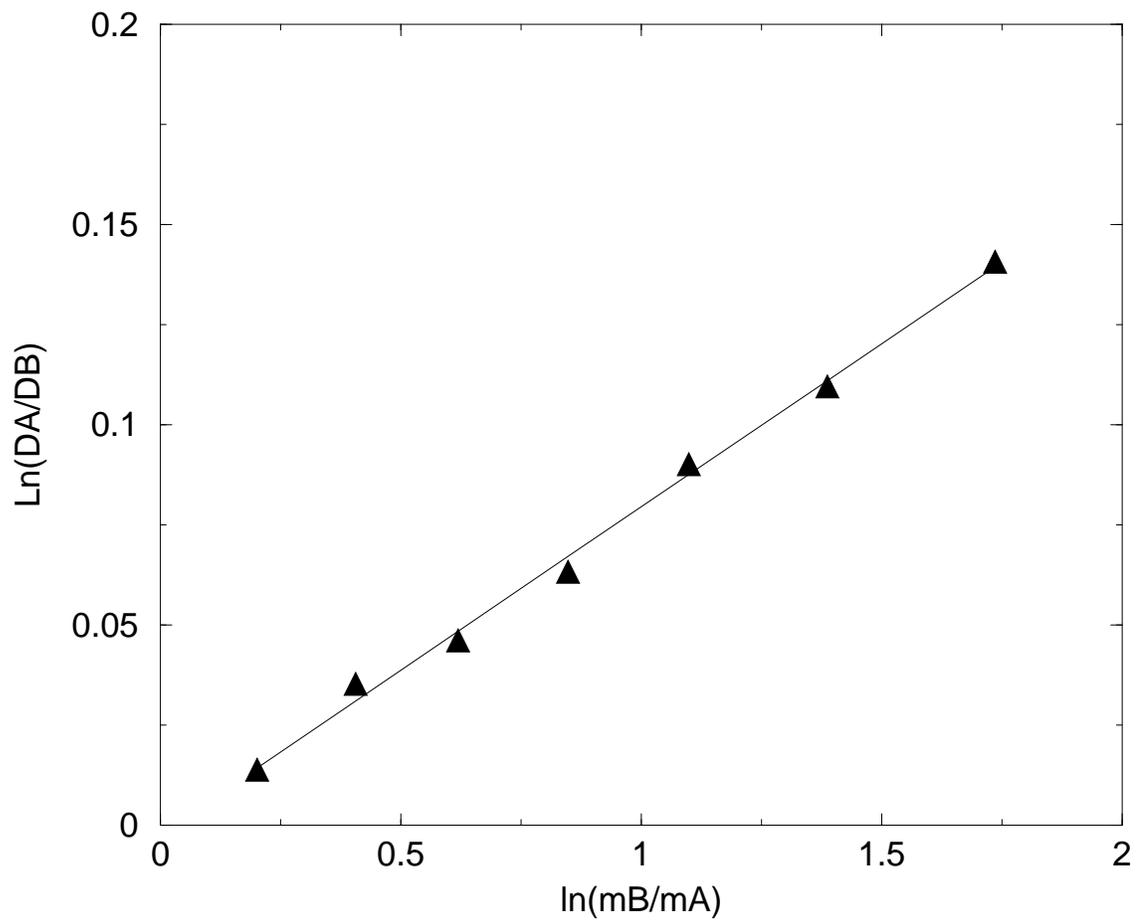}\hfil}
\caption{$\ln(D_A/D_B)$ versus $\ln(m_B/m_A)$. Triangles are results of 
computer simulation and the solid line is the best fit line. $\gamma$ is 
the slope of this line.}
\label{isoteff}
\end{figure}

An additional parameter which could reveal the diffusion mechanism in
isotope mixtures is the isotope effect parameter. It was shown by Parrinello 
{\it et al.} in \cite{Parrinello} that for dilute gases the ratio of the
self-diffusion coefficients of a binary mixture is equal to the reciprocal
of the square root of masses of the two species. However, computer
simulations performed by Bearman and Jolly \cite{Bearman}) show that the
ratio of the self-diffusion coefficients varies with masses as

\begin{equation}
\label{i1}\frac{D_A}{D_B}=\left( \frac{m_B}{m_A}\right) ^\gamma , 
\end{equation}
where $\gamma $ differs from $1/2$ for higher densities and lower
temperatures, when effects of collective motion is important. In Fig. \ref
{isoteff} we plot $\ln (D_A/D_B)$ versus $\ln (m_B/m_A)$. The slope of the
best fit line gives $\gamma =0.081$. This corresponds to well-known
results (for example, therein Ref. \cite{Kluge}). There is another parameter 
that serves
as a quantitative measure of collectivity, the isotope effect, defined as 
\cite{Kluge}

\begin{equation}
\label{i2}E=\frac{D_A/D_B-1}{\sqrt{m_B/m_A}-1}. 
\end{equation}

Large isotope effects are interpreted as single-atom jumps via vacancies 
\cite{Zollmer}, whereas small isotope effects indicate the presence of some 
collective processes. Kluge and Schober in \cite{Kluge} estimated the number
of particles moving cooperatively as

\begin{equation}
\label{i3}N\approx \frac 1E. 
\end{equation}

We find that for our binary mixtures the isotope effect $E$ varies from $%
0.132$ for $\Delta m=0.1$ to $0.109$ for $\Delta m=0.7$. It indicates an
increase in the cooperative motion of particles with an increase in their mass
differences.

\section{Conclusion}

We have considered MD computer simulations of a toy model, a binary isotope
mixture, that allows us to elucidate the effect of having a mass difference on
properties of liquids. It has been shown that redistribution of mass 
($\Delta m$) among the two species, while keeping the rest of parameters 
fixed, does not affect static properties (the pair distribution function 
$g(r)$) both for the system and separately for each species $A$ and $B$ and
these are independent of $\Delta m$). At the same time, all dynamic properties 
(the diffusion coefficient, velocity autocorrelation function, non-Gaussian and
isotope effect parameters) are affected and exhibit $\Delta m$ dependence,
which demonstrates the pure dynamical nature of mass heterogeneity.

Moreover, the $\Delta m$ dependence exhibits some ``critical'' values such as 
in the behavior of the negative part of the normalized velocity autocorrelation
function $\psi (t)$ (see Fig. \ref{vacf}) which at $\Delta m=0.5$ changes
the initial tendency of decreasing magnitude of first minimum with increase
of $\Delta m$ to the opposite, i.e. a drastic increase in $\psi (t)$. 
Such phenomenon, we believe, deserves further investigation and analysis.

Many real systems have mass heterogeneities but, of course, not pure
as in our model. Nevertheless, even the hidden and probably not very strong
effect of having a mass difference compared to other (size, interaction, etc.) 
has to create a pure dynamical contribution. As demonstrated by our simple 
model, this pure mass effect does not affect the static properties of a system.
In particular, this observation leads us to question the validity of some
universal empirical formulae based on the conjecture that ``atomic diffusion is
an entirely geometrical phenomenon'' \cite{Dzugutov}. The author of \cite
{Dzugutov} claimed that the relationship between $D$ and $g(r)$ is governed
by following equations

\begin{equation}
\label{10}D^{*}=0.049e^{S_2},
\end{equation}
where $D^{*}$ is the diffusion coefficient in the dimensionless form $%
D^{*}=D\Gamma _E^{-1}\sigma ^{-2}$, $\Gamma _E$ is given by

\begin{equation}
\label{11}\Gamma _E=4\sigma ^2g(\sigma )\rho \sqrt{\pi kT/m},
\end{equation}
where $\sigma $ is a position of first maximum of $g(r)$ and $S_2$ is the
two-particle approximation of the reduced excess entropy 

\begin{equation}
\label{12}S_2=-2\pi \rho \int_0^\infty \left\{ g(r)\ln \left[ g(r)\right]
-\left[ g(r)-1\right] \right\} r^2dr.
\end{equation}

According \cite{Dzugutov}, Eqs. (\ref{10},\ref{11},\ref{12}) are universal for
equilibrium condensed atomic systems, both liquid and solid, regardless of
the structures, interatomic interaction potential or the microscopic
dynamical mechanisms involved and also valid for multicomponent systems with 
$\Gamma _E$, $\sigma $ and $S_2$ corresponding to each type of constituent 
atoms. 

This universality contradicts results of our simulation because
Eqs. (\ref{10},\ref{11},\ref{12}) for the considered binary mixture 
give Eq. (\ref{i1}) 
with $\gamma =0.5,$ whereas from Fig. \ref{isoteff} the slope of the
line of best fit gives $\gamma =0.081$.

For our system the observed difference in the diffusion coefficients of two
species is an entirely nongeometrical, nonstatic phenomenon. 

\section{Acknowledgments}

We thank S. Consta and D.G.C. McKeon for valuable comments and reading the 
manuscript.

\newpage
\begin{table}[h]
\begin{center}
\begin{tabular}{|r|r|r|r|r|r|r|r|}
\hline
$m_A$ & $m_B$ & $D_{A(E)}$ & $D_{A(K)}$ & $D_{B(E)}$ & 
$D_{B(K)}$ & $D_{(E)}$ & $D_{(K)}$ \\ 
\hline
1 & 1 & 5.4979 & 5.4749 & 5.4979 & 5.4749 & 5.4979 & 5.4749 \\ 
\hline
0.9 & 1.1 & 5.5861 & 5.4866 & 5.4592 & 5.7079 & 5.5355 & 5.6564 \\ 
\hline
0.8 & 1.2 & 5.6677 & 5.6369 & 5.4710 & 5.4368 & 5.5797 & 5.4842 \\ 
\hline
0.7 & 1.3 & 5.7393 & 5.7635 & 5.4802 & 5.5991 & 5.6108 & 5.6169 \\ 
\hline
0.6 & 1.4 & 5.9216 & 5.8336 & 5.5586 & 5.5582 & 5.7045 & 5.7818 \\ 
\hline
0.5 & 1.5 & 6.1117 & 6.0612 & 5.5852 & 5.6850 & 5.8550 & 5.9024 \\ 
\hline
0.4 & 1.6 & 6.3923 & 6.6003 & 5.7286 & 5.8319 & 6.0542 & 6.1483 \\ 
\hline
0.3 & 1.7 & 6.7438 & 6.5358 & 5.8589 & 6.0439 & 6.3819 & 6.4645 \\ 
\hline
\end{tabular}
\caption{\label{tab1} The dependence of the diffusion coefficient on
masses of species. All $D$ are multiplied by $10^{-2}$
(in scaled units).}
\end{center}
\end{table}

\newpage

\end{document}